\documentclass[a4paper,12pt]{article}
\usepackage{mystyle} 

\usepackage[T1]{fontenc} 
\usepackage{tikz,caption,subcaption} 
\usetikzlibrary{decorations.markings}

\usepackage{epsfig,euscript,xspace, color}
\usepackage{mathtools}

\def\bea{\begin{eqnarray}}
\def\eea{\end{eqnarray}}
\def\be{\begin{equation}}
\def\ee{\end{equation}}

 \newcommand{\pd}{\partial}
\newcommand{\prd}[2]{\frac{\partial #1}{\partial #2}}

\newcommand{\td}{\text{d}}
\newcommand{\trd}[2]{\frac{\text{d} #1}{\text{d} #2}}
\newcommand{\mc}[1]{\mathcal{#1}}
\newcommand{\grad}{\vec{\nabla}}
\newcommand{\Exp}[1]{\text{e}^{#1}}
\newcommand{\Log}[1]{\text{ ln}\left(#1\right)}

\newcommand{\Cos}[1]{\text{ cos}\left(#1\right)}

\newcommand{\no}[1]{:#1:}
\newcommand{\bra}{\langle}
\newcommand{\ket}{\rangle}
\newcommand{\cross}{\times}

\newcommand\AdS{$AdS_3$\xspace}

\usetikzlibrary{arrows,snakes}

\definecolor{lightblue}{rgb}{.1,.4,.5}
\definecolor{brown1}{rgb}{.64,.43,.138}

\usepackage{mathrsfs}  

\title{\boldmath Sine-Gordon Theory : Entanglement entropy and holography}

\author[a,b,c]{Pinaki Banerjee}
\author[a,b]{\! Atanu Bhatta}
\author[a,b]{\! and B. Sathiapalan}

\affiliation[a]{\vspace{.3cm} Institute of Mathematical Sciences\\ CIT Campus, Taramani \\ Chennai - 600 113, India \\ \vspace{-.3cm}
}
\affiliation[b]{Homi Bhabha National Institute\\ Training School Complex, Anushakti Nagar \\ Mumbai 400085, India\\  \vspace{-.3cm}
}
\affiliation[c]{
	International Centre for Theoretical Sciences, \\
	Tata Institute of Fundamental Research, \\
	Shivakote, Bengaluru 560089, India\\  \vspace{.1cm}
}

\emailAdd{pinakib@imsc.res.in}
\emailAdd{batanu@imsc.res.in}
\emailAdd{bala@imsc.res.in}

\abstract{We compute change in entanglement entropy for a single interval in $1+1$ dimensional
sine-Gordon model perturbatively in the coupling. The sine-Gordon perturbation can be thought of as 
deformation of the free CFT by a primary operator with dimension $\Delta$. In an independent 
computation we calculate holographic entanglement entropy for that interval from three dimensional 
bulk AdS which has a massive scalar with its mass satisfying $m^2= \Delta(\Delta-2)$. We show that 
the two results match for near-marginal perturbations upto leading order in the coupling.}

\begin{document} 
\hspace{12cm} IMSc/2016/10/05

\maketitle
\flushbottom

\section{Introduction}
  Entanglement entropy (EE) \cite{Bombelli:1986rw,Srednicki:1993im,Holzhey:1994we,
  Calabrese:2004eu} is a measure of entanglement for a quantum state. Besides the conceptual
  importance of understanding it because of its purely quantum nature, it has been proposed as 
  a practical way to distinguish between phases in some condensed matter systems that cannot be 
  distinguished by any local order parameter \cite{Kitaev:2005dm,Levin:2006zz,Vidal:2002rm}.

 There is a simple elegant geometric prescription for computing  the entanglement entropy of 
 a system using holographic techniques \cite{Ryu:2006bv,Ryu:2006ef,Hubeny:2007xt}. 
 As a result, a lot of work has been done in evaluating this quantity on the bulk side. 
 The corresponding calculation for conformal field theory (CFT) in the boundary has also 
 received a lot of attention and wherever both techniques are applicable there is a perfect 
 match (see for example \cite{Hartman:2013mia,Faulkner:2013yia, Caputa:2014vaa, 
 Asplund:2014coa,  Nozaki:2014hna,Banerjee:2016qca,Datta:2014ska,Datta:2014uxa,
 Datta:2014zpa} and references therein). There are also derivations of the holographic
 prescription in different situations viz. for spherical entangling regions \cite{Casini:2011kv}, 
 for time independent scenario \cite{Lewkowycz:2013nqa} and recently for the covariant
 version of the conjecture \cite{Dong:2016hjy}.
  
  A CFT describes a fixed point under renormalization group (RG) of a more general 
  field theory. Typically in a field theory there is an RG flow from a ultra-violate (UV)
  fixed point to an infra-red (IR) fixed point and the field theory  lives on the 
  trajectory in between the two fixed points. It is only for a very special choice of 
  parameters that it is exactly at one of the fixed points and then it is expected to be 
  conformally invariant. One expects the holographic correspondence to be valid not only
  at both fixed points but also along the entire trajectory. In the bulk this flow has
  been termed ``Holographic RG'' \cite{Akhmedov:1998vf,deBoer:1999tgo,Balasubramanian:1999jd,deHaro:2000vlm,
Skenderis:2002wp,Bianchi:2001kw,Peet:1998wn,Witten:2001ua,Heemskerk:2010hk,
Faulkner:2010jy,Lizana:2015hqb,Sathiapalan:2017frk} although it is far from clear what
exactly the connection is with the usual Wilsonian RG. In particular the details of the 
regulator or the coarse graining procedure have not been satisfactorily worked out 
\cite{Heemskerk:2010hk, Lizana:2015hqb,Sathiapalan:2017frk}. 

In order to understand these issues better it is useful to calculate physical quantities
away from the fixed point and try to check the holographic correspondence along the RG 
trajectory. Entanglement entropy is one such useful physical quantity and in this paper 
we check this correspondence at points on the trajectory away from, but close to, the 
fixed point of a field theory (see also \cite{Rosenhaus:2014zza,Rosenhaus:2014nha, 
Lewkowycz:2012qr,Herzog:2013py}). 
In order to keep the computations as simple as possible, we consider a field theory 
in 1+1 dimension and its holographic dual in 2+1 dimension.  Also to make things concrete we 
consider a specific field theory - the sine-Gordon theory\footnote{See \cite{Cardy:2007mb,CastroAlvaredo:2008kh,
Doyon:2008vu, CastroAlvaredo:2009ub, Levi:2013sba} for related  works in EE for 
sine-Gordon and other integrable models.}.
\begin{align}
\label{eq:bare_action}
\mc{A} = \int \td^2x \left[\frac{1}{2} ( \grad \phi )^2 -\frac{\lambda_0}{\beta^2 a^2} \text{cos}(\beta \phi)\right]
\end{align}

This is a very interesting and non trivial field theory in its own right. In particular it
is related to the XY model and has the well known Kosterlitz-Thouless phase transition 
\cite{Kosterlitz:1973xp,Kosterlitz:1974sm,Frohlich:1981yn}. The complete solution
of the model \emph{i.e,} the full description of its exact scattering matrix and particle
spectrum were given in \cite{Zamolodchikov:1976uc}. Different integrable models have also been studied away from criticality using conformal perturbation theory (see e.g, \cite{Zamolodchikov:1989cf, Zamolodchikov:1990bk}). In the context of string theory this action is also a world sheet description of a particular tachyonic background. This
has been exploited in obtaining the equations of motion for the tachyon - which are 
generalizations of the sine-Gordon $\beta$- functions \cite{Das:1986cz}. 

  \begin{center}

  	\tikzset{middlearrow/.style={
  			decoration={markings,
  				mark= at position 0.5 with {\arrow{#1}} ,
  			},
  			postaction={decorate}
  		}
  	}

  	\begin{tikzpicture}[scale=0.75]
  	
  	\draw [->, line width= 1 mm] (0,0) --(5,0);
  	\node [below right] at (5,-0.2) {\large{$\delta$}};
  	\node [below, black] at (2.5,-0.2) {Fixed line};
  	\draw [-, thick, ->] (0,0) --(0,5);
  	\node [below right] at (0.2,5.4) {\large{$\lambda$}};
  	\draw [-, thick, <-] (-5,0) --(0,0);
  	\draw [blue,-,thick,  middlearrow={<}] (0,0) --(4.2,4.2);
  	\node [below right] at (4.2,4.2) {\textcolor{black}{S$_1$}};
  	\draw [blue,-,thick, middlearrow={>}] (0,0) --(-4.2,4.2);
  	\node [below left] at (-4.2,4.2) {\textcolor{black}{S$_2$}};

  	\draw [blue,thick,domain=0:2, middlearrow={<}] plot ({1.*cosh(\x)}, {1.*sinh(\x)});
  	\draw [blue,thick,domain=0:1.58, middlearrow={<}] plot ({1.5*cosh(\x)}, {1.5*sinh(\x)});
  	\draw [blue,thick,domain=0:1.27, middlearrow={<}] plot ({2*cosh(\x)}, {2*sinh(\x)});

  	\draw [blue,thick,domain=0:2, middlearrow={>}] plot ({-1.*cosh(\x)}, {1.*sinh(\x)});
  	\draw [blue,thick,domain=0:1.58, middlearrow={>}] plot ({-1.5*cosh(\x)}, {1.5*sinh(\x)});
  	\draw [blue,thick,domain=0:1.27, middlearrow={>}] plot ({-2*cosh(\x)}, {2*sinh(\x)});

  	\draw [blue,thick,domain=0:2, middlearrow={<}] plot ({1.*sinh(\x)}, {1.*cosh(\x)});
  	\draw [blue,thick,domain=0:1.58, middlearrow={<}] plot ({1.5*sinh(\x)}, {1.5*cosh(\x)});
  	\draw [blue,thick,domain=0:1.27, middlearrow={<}] plot ({2.*sinh(\x)}, {2.*cosh(\x)});

  	\draw [blue,thick,domain=0:2, middlearrow={>}] plot ({-1.*sinh(\x)}, {1.*cosh(\x)});
  	\draw [blue,thick,domain=0:1.58, middlearrow={>}] plot ({-1.5*sinh(\x)}, {1.5*cosh(\x)});
  	\draw [blue,thick,domain=0:1.27, middlearrow={>}] plot ({-2.*sinh(\x)}, {2.*cosh(\x)});

  	\node at  (3,1) {\textcolor{red}{I}};
  	\node at  (-3,1) {\textcolor{red}{III}};
  	\node at  (0.3,3) {\textcolor{red}{II}};
  	
  	\end{tikzpicture}
  	
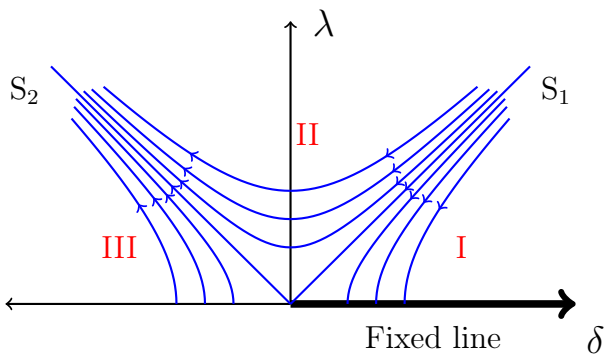
\captionof{figure}{\emph{RG flow of sine-Gordon theory. There are three different regions 
  	separated by the flow lines S$_1$ and S$_2$. }} \label{Fig:flow_SG}
  \end{center}

\vspace{0.5cm}

In this paper we calculate the change in entanglement entropy along the flow of the coupling 
$\lambda$. The RG  flow has three regions corresponding to three different phases \cite{Amit:1979ab}. 
See the RG flow diagram for sine-Gordon theory in Fig.~\ref{Fig:flow_SG}. In region I the theory 
flows in the IR to a free theory. This is where the cosine perturbation is irrelevant. The theory
is on the critical surface and the correlation length is infinite. In region II the correlation length
is finite and depends on which RG trajectory the theory is on, which in turn is determined in terms of 
the parameters of the theory. Because the theory flows towards strong coupling in region II and III and
to weak coupling in region I, our perturbative calculation of entanglement entropy is easier to justify in region I.

We use a technique introduced by  Holzhey, Larsen and Wilczek (HLW) \cite{Holzhey:1994we} involving the trace anomaly.
To calculate the EE corresponding to a finite interval, HLW \cite{Holzhey:1994we} map this interval to the
infinite half line using a conformal transformation and proceed to calculate the entanglement entropy of this 
system by extracting the dependence of the EE on $\ln a$ ($a$ is the short distance cutoff) using the trace anomaly. 
They then argue that because the theory is conformal, the only other scale is the length of the interval and therefore
the dependence is in fact $\ln {l\over a}$. The only subtlety in using the same technique for this perturbed 
theory is that the theory is not conformal anymore. Therefore one might think the conformal transformation used 
by HLW \cite{Holzhey:1994we} to map the finite interval to half line cannot be applied for this case. However we 
argue in section \ref{sec:EE}, if we are interested in change in EE only up to \emph{leading order} in $\Delta-2$,
we can still use this map. And following HLW we can extract the dependence on $\ln a$.

If the theory is in region II or III, it has a finite correlation length - $\xi$ and is the only other scale
in theory (assuming $l$ is infinite). Thus we expect a dependence $\ln {\xi\over a}$. The correlation length
in region II has been estimated in  \cite{Amit:1979ab}. Defined as the scale where the coupling constant 
$\delta $ $ (=\beta^2/8\pi - 1)$ is of O(1) they obtain $\xi = a e^{4\pi \over \lambda_0}$ where $\lambda_0$
is the intercept of the RG trajectory on the y-axis - which corresponds to the line ${\beta^2\over 4\pi}-2=0$, 
when the cosine is a marginal perturbation near the trivial fixed point. Thus $\ln {\xi\over a}={4\pi \over \lambda_0}$. 
While this quantity is an RG invariant,  the coupling constants flow along the trajectory. And there is 
some scale dependence in the entanglement entropy. Defining entropy involves coarse graining and a scale 
dependence is not unexpected. This scale dependence is similar to that in the $c$-function defined by 
Zamolodchikov \cite{Zamolodchikov:1986gt}. Here a distance scale enters - the distance between the two operators
in the two point function of components of the energy momentum tensor used to define $C(r)$. It becomes unambiguous 
only at fixed points.

However if one works with finite subsystem of length $l$, near criticality\footnote{We will see in section \ref{sec:EE} that relevant perturbations can be treated in our perturbative method as long as it is very close to marginal, such that $\xi \gg l$. But going a finite distance in RG flow is beyond the scope of this paper and is for  future research.} even for $\Delta < 2$ the correlation length $\xi \gg l$ and the EE then scales as $\ln {l\over a}$. Both finite and semi-infinite subsystems have been considered in this paper. The finite $l$ case is compared with the holographic result. 

In region I, the cosine perturbation is marginally irrelevant and we are on the critical surface
and $\xi=\infty$. The only scale that can possibly enter in the logarithm is the IR cutoff, which we can 
take to be $l$ and we can expect again that the dependence will be $\ln {l\over a}$. In this case also one 
should interpret
the coupling as flowing from the UV scale $a$ to the IR scale $l$.  The coupling in the IR limit 
is zero and the coupling in the expression for entanglement entropy should be interpreted as the value at the UV scale.

The entanglement entropy is also related to the central charge of the theory by $S_{E}={c\over 3}\ln {l\over a}$. 
This relation is true for exactly conformal field theories.   One may expect that, at least to lowest order,
this continues to be true  along the RG flow if we use a suitably defined 
``central charge function\footnote{ For more details on c-function and how it is related to holographic EE 
see \cite{Myers:2010tj, Myers:2010xs}.}'' \cite{Zamolodchikov:1986gt,Bloete:1986qm, Komargodski:2011vj,
Komargodski:2011xv}. At higher orders there should be ambiguities related to the choice of flow equation 
which we know is far from unique - this non uniqueness has to do with the choice of regulation or coarse
graining that is adopted along the flow. In any case we test this by considering the change in central charge. 
This has been calculated using the exact renormalization group (ERG)  and also other methods \cite{Oak:2017trw}. 
We  find that the answers match with that obtained by the other two methods. As mentioned above, at lowest order 
this match is not unexpected.

There is one noteworthy feature about this match. If the change in entanglement entropy is related 
to the change in central charge, then it is clear that it must decrease along the RG flow towards the IR.
Thus if the perturbation is relevant then
$\lambda>0$ is the IR end of the flow and $\lambda=0$ is the UV end of the flow. If the perturbation 
is irrelevant the opposite holds. Thus the sign of the change in EE between $\lambda>0$ and $\lambda=0$ 
should reflect this, even though the lowest order answer depends on $\lambda^2$. It is reassuring that 
the final answer does satisfy this requirement.

The rest of the paper is organised as follows. In section \ref{sec:EE} we compute the EE for sine-Gordon
theory in leading order in coupling using 2D field theoretic techniques. We calculate the same quantity 
in section \ref{sec:HEE} but from bulk using holography. Section \ref{sec:conclusions} contains the summary
and interpretation of our results and some future directions.

  \section{\label{sec:EE}Entanglement entropy from 2D field theory}
 Sine-Gordon theory \cite{Coleman:1974bu,Mandelstam:1975hb} on a 2-dimensional Euclidean space 
 is described by the action,
  \begin{equation}
  \label{eq:bare_action}
   \mc{A} = \int \td^2x \left[\frac{1}{2} ( \grad \phi )^2 -\frac{\lambda_0}{\beta^2 a^2} \text{cos}(\beta \phi)\right]
  \end{equation}
  $\lambda_0$ represents bare coupling and $a$ is the UV cutoff. 
  In the complex plane the action (\ref{eq:bare_action}) becomes,
  \begin{align}
  \label{eq:bare_complex_action}
   \mc{A} &= \int \td^2z \left[ \pd_z \phi \,\pd_{\bar{z}} \phi -\frac{1}{2}\frac{\lambda_0}{\beta^2 a^2} \text{cos}(\beta \phi(z))\right] \nonumber \\
          &\equiv \mc{A}_0 + \mc{A}_{SG}
  \end{align}

  Superficially the interaction sine term looks like a relevant interaction because the scalar
  field in 2 dimensions is dimensionless. Nevertheless at the quantum level it has a well defined 
  anomalous dimension and can be relevant, marginal or irrelevant. At leading order this is just 
  determined by the parameter $\beta$ since the dimension of the operator is $\beta^2\over 4\pi$.
  
   For a quantum mechanical system with many degrees of freedom 
  the density matrix is given by,
  \begin{equation}
   \rho_{tot} = |\Psi\rangle \langle \Psi |
  \end{equation}

  where $ |\Psi\rangle $ is the state vector of the system. 
  Now we divide the total system into two subsystems $A$ and $B$.
  
   \vspace{1cm}
  \begin{center}
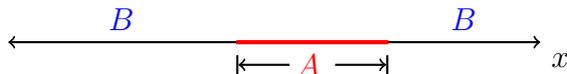

    \begin{tikzpicture}
\draw [<->, thick] (-3,0) --(4,0);
\node [below right] at (4,0) {$x$};
\draw [ultra thick,red] (0,0) --(2,0);
\draw [|<-, thick] (0,-0.3) --(0.7,-0.3);
\draw [->|, thick] (1.3,-0.3) --(2,-0.3);
\node [left] at (1.25,-0.3) {\textcolor{red}{$A$}};
\node [left] at (-1.2,.3) {\textcolor{blue}{$B$}};
\node [left] at (3.3,.3) {\textcolor{blue}{$B$}};

\end{tikzpicture}
\captionof{figure}{\emph{Sub-systems $A$ and $B$ of an infinite 1D system}}
   \label{Intervals}

  \end{center}

    \vspace{0.5cm}

  Tracing out the degrees of freedom of $B$, we are left with the reduced density matrix,
  \begin{equation}
   \rho_A = \text{Tr}_B (\rho_{tot})
  \end{equation}
  which describes the remaining degrees of freedom in $A$. The entanglement entropy across $\pd A$ is then
  given by von Neumann entropy of $\rho_A$,
  \begin{equation}
   S_{\pd A} = -\text{Tr}_A (\rho_A \text{ log }\rho_A).
  \end{equation}
  We are interested in computing entanglement entropy for the sine-Gordon theory. For that purpose
  we consider an interval of length $l$ (see Fig. \ref{fig:system}) and compute its EE. This is a
  measure of how much this interval is quantum mechanically entangled to the rest of the system. 
  
But it is well known that the EE computed for this system will diverge as there is no UV cutoff. 
To regularize that divergence let's introduce two UV cutoffs namely $a_1$ and $a_2$ at the end points 
of the subsystem $A$. For simplicity we take $a_1 = a_2 = a$ (see Fig.~\ref{fig:system}).

\vspace{0.5cm}

  \begin{center}
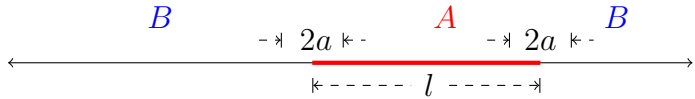

    \begin{tikzpicture}
\draw [<->, ] (-4,0) --(5,0);
\draw [ultra thick,red] (0,0) --(3,0);
\draw [|<-, dashed] (0,-0.3) --(1.3,-0.3);
\draw [->|, dashed] (1.8,-0.3) --(3,-0.3);
\node [left] at (1.75,-0.3) {$l$};

\node  at (1.75,0.6) {\textcolor{red}{$A$}};
\node  at (4,0.6) {\textcolor{blue}{$B$}};
\node  at (-2,0.6) {\textcolor{blue}{$B$}};

\draw [|<-, dashed] (0.4,0.3) --(0.7,0.3);
\draw [->|, dashed] (-.7,0.3) --(-.4,0.3);
\draw [->|, dashed] (2.3,0.3) --(2.6,0.3);
\draw [|<-, dashed] (3.4,0.3) --(3.7,0.3);
\node  at (3.0,0.3) {$2 a$};
\node  at (0.05,0.3) {$2 a$};
\end{tikzpicture}

\captionof{figure}{\emph{The sub-system $A$ with the UV cutoff $a$ at its boundary points.}}
   \label{fig:system}
  \end{center}

    \vspace{1cm}

HLW compute EE for such a system but for a conformal field theory in its ground state. 
In their paper \cite{Holzhey:1994we} HLW introduce an IR cutoff and then map the subsystem $A$ to a
half line by a standard conformal mapping. Due to that special transformation the IR cutoff decouples
and only available length scales are the subsystem size $l$ and the UV cutoff $a$. Thus $l\over a$ is 
the only dimensionless quantity for that problem. They probe the UV sensitivity of the partition function
to obtain the famous logarithmically divergent EE for the critical system : $S_{E}={c\over 3}\ln {l\over a}$.
We start with a system whose size is very large\footnote{The reason behind taking (semi-) infinitely 
large system size is lack of conformal invariance. Considering a finite interval amounts to adding a 
scale to the problem. Therefore one would naively think the scale invariance will be broken and the 
conformal transformation \eqref{half-line-map} of HLW that maps finite interval to a half line will not
keep the action form invariant. Although this is generically true, we explicitly show later in this section 
that the effect doesn't show up at leading order in $(\Delta-2)$.} and can itself be considered as a half line.
At the end of this section we shall show that our results holds for any finite system with arbitrary size $l$ to 
the linear order in $\delta \ (i.e, \ \text{in} \ \Delta -2$).

\vspace{0.5cm}
  \begin{center}
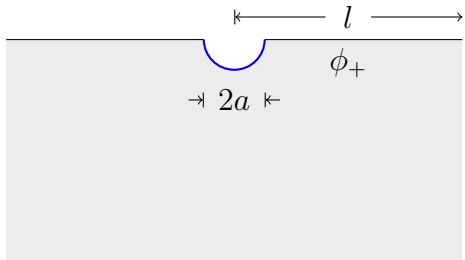

    \begin{tikzpicture} [scale = 1]
\fill[gray!15]  ( -3,-3) |- (-0.4,0) arc (-180:0:4mm) -| (3,-3);
\draw ( -3,0) -- (-0.4,0)  (0.4,0) -- ( 3.0,0);
\draw [blue,thick] (-0.4,0) arc  (-180:0:4mm);

\draw [|<-] (0,0.3) --(1.2,0.3);
\draw [->] (1.8,0.3) --(3,0.3);
\node at (1.5,0.3) {$l$};

\node at (1.5,- 0.3) {$\phi_+$};

\draw [->|] (0.6,-0.8) --(0.4,-0.8);
\draw [|<-] (-.4,-0.8) --(-.6,-0.8);
\node  at (0.,-0.8) {$2 a$};
   \end{tikzpicture}

\captionof{figure}{\emph{Our sub-system is a semi-infinite line. The strip represents the ground 
state wave functional for the whole system.}} \label{fig:strip_1}
  \end{center}

\vspace{0.2cm}

Fig.~\ref{fig:strip_1} represents the ground state wave-functional for the $1+1$ dimensional 
field theory which is obtained by path-integrating the field from $t = -\infty$ to $t = 0$ in 
the Euclidean formalism. The values of the field $\phi_+$ at the boundary depends on the spatial
coordinate. The total density matrix is given by two copies of the wave functional.
\begin{align}
[\rho]_{\phi_+,\phi_-} = \Psi[\phi_+] \bar{\Psi}[\phi_-]
\end{align}

The complex conjugate one can be obtained by path-integrating from $t = +\infty$ to $t = 0$. 
To obtain the reduced density matrix we integrate over the subsystem $B$ (see Fig.~\ref{fig:density_matrix}) 
which is equivalent to sewing the two sheets along $B$.

\vspace{0.4cm}

  \begin{center}
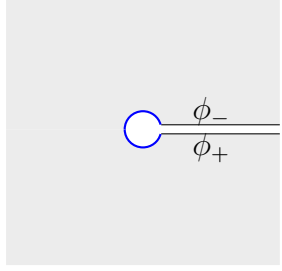

    \begin{tikzpicture} [scale = 0.6]
\fill[gray!15]  ( -3,-3) |- (-0.4,0) arc (-180:-15:4mm) -| (3,-3);
\draw   (0.4,-0.1) -- ( 3.0,-0.1);
\draw [blue,thick] (-0.4,0) arc  (-180:-15:4mm);

\node at (1.5,- 0.4) {$\phi_+$};

\fill[gray!15]  ( -3,3) |- (-0.4,0) arc (180:15:4mm) -| (3,3);
\draw   (0.4,.1) -- ( 3.0,0.1);
\draw [blue,thick] (-0.4,0) arc  (180:15:4mm);
\node at (1.5, 0.4) {$\phi_-$};
   \end{tikzpicture}
\captionof{figure}{\emph{Pictorial representation of the reduced density matrix. The boundary
values of the field $\phi_+,\phi_-$ are the entries of the density matrix
$[\rho_A]_{\phi_+,\phi_-}$. }} \label{fig:density_matrix}
  \end{center}
One can compute the EE for subsystem $A$ using replica trick. 
  \begin{align}\label{replica}
  {S_{\pd A}}= \left(1-n\frac{\text{d}}{\td n}\right)\text{ln}\, \mathcal{Z}(n)\Bigg{|}_{n=1}
  \end{align}     
  where $ \mathcal{Z}(n) =\mathcal{Z}(1)^n tr(\rho_A^n) $. This $tr(\rho_A^n)$ can be 
  computed by introducing $n$ such sheets and sewing them in a particular manner 
  (see Fig~\ref{fig:Z(3)} for $n=3$ case). The topology of the replica surface\footnote{Ref. \cite{Cardy:2007mb} also computes EE in integrable 1+1 theories with large sub-system size away from criticality by computing correlation functions of branch point twist fields which are symmetry fields associated to the cyclic permutation symmetry of the replica theory. In particular in \cite{Doyon:2008vu} the EE of an interval in sine-Gordon model was studied using similar method (for other applications of this technique see \cite{Bianchini:2016mra, Bianchini:2015uea,CastroAlvaredo:2008pf,CastroAlvaredo:2011zs}.).} \cite{Calabrese:2004eu, Cardy:2007mb} becomes a cone with angular circumference $2 \pi n$. 

\vspace*{-3.7cm}  
  
  \begin{center}
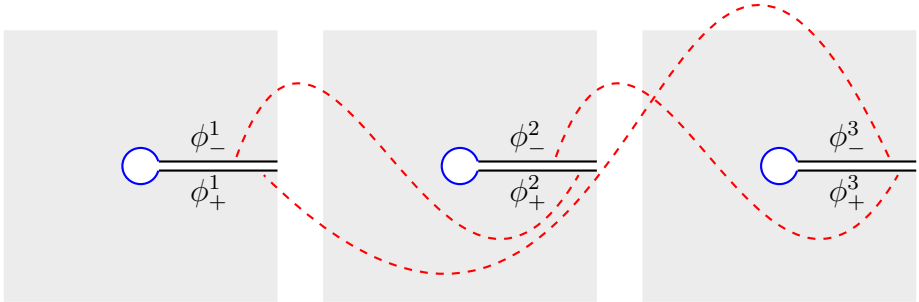

    \begin{tikzpicture} [scale = 0.6]
\fill[gray!15]  ( -3,-3) |- (-0.4,0) arc (-180:-15:4mm) -| (3,-3);
\draw  [thick] (0.4,-0.1) -- ( 3.0,-0.1);
\draw [blue,thick] (-0.4,0) arc  (-180:-15:4mm);

\node at (1.5,- 0.6) {$\phi^1_+$};

\fill[gray!15]  ( -3,3) |- (-0.4,0) arc (180:15:4mm) -| (3,3);
\draw   [thick] (0.4,.1) -- ( 3.0,0.1);
\draw [blue,thick] (-0.4,0) arc  (180:15:4mm);
\node at (1.5, 0.6) {$\phi^1_-$};

\fill[gray!15]  ( 4,-3) |- (6.6,0) arc (-180:-15:4mm) -| (10,-3);
\draw  [thick] (7.4,-0.1) -- ( 10.0,-0.1);
\draw [blue,thick] (6.6,0) arc  (-180:-15:4mm);

\node at (8.5,- 0.6) {$\phi^2_+$};

\fill[gray!15]  ( 4,3) |- (6.6,0) arc (180:15:4mm) -| (10,3);
\draw  [thick] (7.4,.1) -- ( 10.0,0.1);
\draw [blue,thick] (6.6,0) arc  (180:15:4mm);
\node at (8.5, 0.6) {$\phi^2_-$};

\fill[gray!15]  ( 11,-3) |- (13.6,0) arc (-180:-15:4mm) -| (17,-3);
\draw   [thick](14.4,-0.1) -- ( 17.0,-0.1);
\draw [blue,thick] (13.6,0) arc  (-180:-15:4mm);

\node at (15.5,- 0.6) {$\phi^3_+$};

\fill[gray!15]  ( 11,3) |- (13.6,0) arc (180:15:4mm) -| (17,3);
\draw  [thick] (14.4,.1) -- ( 17.0,0.1);
\draw [blue,thick] (13.6,0) arc  (180:15:4mm);
\node at (15.5, 0.6) {$\phi^3_-$};

\draw [red,dashed,thick]  (2.1, 0.2)  to[out=70,in=-120, distance=6cm ] (9.6,- 0.2);

\draw [red,dashed,thick]  (9.1, 0.2)  to[out=70,in=-120, distance=6cm ] (16.6,- 0.2);

\draw [red,dashed,thick]   (16.4, 0.2) to[out=115,in=-45, distance=12cm ] (2.7, -0.2);

   \end{tikzpicture}

\vspace{-2.5cm}  

\captionof{figure}{\emph{$tr(\rho_A^3) = [\rho_A]_{\phi^1_+,\phi^1_-} [\rho_A]_{\phi^2_+,\phi^2_-} [\rho_A]_{\phi^3_+,\phi^3_-}$ where we identify $\phi^1_- \sim \phi^2_+$, $\phi^2_- \sim \phi^3_+$ and $\phi^3_- \sim \phi^1_+$. }} \label{fig:Z(3)}
  \end{center}

The partition function for a field theory on a manifold $M$ is given by,
  \begin{equation}
   \mc{Z} = \int_M \mc{D}\phi \, \Exp{-\mc{A}[\phi]}
  \end{equation}
  Under a global dilatation, $x^\mu \to (1 - 2 \,\alpha) x^\mu$, the action changes as,
  \begin{equation}
   \delta \mc{A} = - 2 \alpha \int_M \td^2x \sqrt{g} \ T^\mu_\mu
  \end{equation}
  therefore the change in partition function,
  \begin{align}
   \delta \text{ ln}\,\mc{Z} &= - \int_M \mc{D} \phi \, (\delta \mc{A}) \Exp{-\mc{A}[\phi]} \nonumber \\
                 &= - \bra \delta \mc{A} \ket \nonumber                
  \end{align}
  where we have considered that the partition function is normalized. Under the dilation the cutoff 
  changes as $a \to (1 + 2 \alpha) a $. With $2 \alpha = \delta a / a$,
    
    \begin{equation}
  \label{eq:ZT_connection}
     \prd{\text{ ln}\,\mc{Z}}{\Log{a}} =  \int_M \td^2x \, \sqrt{g} \, \bra T^\mu_\mu \ket.
  \end{equation} 
From \eqref{replica} and \eqref{eq:ZT_connection} one obtains \cite{Holzhey:1994we} 

    \begin{equation}
   \label{eq:master_formula}
   \prd{\,S_{\pd A}}{\,\text{ln} a}= \left(1-n\frac{\text{d}}{\td n}\right)_{n=1} \prd{\,\text{ln}\mathcal{Z}(n)}{\,\text{ln}a} = \left(1-n\frac{\text{d}}{\td n}\right)_{n=1} \,\int_{M_n} \sqrt{g}\, \td^2 x \, \langle T^\mu_\mu \rangle
  \end{equation}
  
Using this formula, Holzhey et al obtained the famous result $S_{EE}= {c\over 3} \ln {l\over a}$. 
We will compute the change $\Delta S_{EE}$ caused by the addition of the cosine perturbation, 
which changes $\bra T^\mu_\mu \ket$. 
Note that our formula \eqref{eq:master_formula} has different normalization factor compared 
to HLW \cite{Holzhey:1994we}. This is due to different conventions of defining the energy-momentum 
tensor $T_{\mu \nu}$. According to their convention, 
$T_{\mu \nu} = \frac{1}{4\pi} \frac{1}{\sqrt{g}}\frac{\delta S}{\delta g^{\mu \nu}}$,
whereas $T_{\mu \nu} = \frac{1}{\sqrt{g}}\frac{\delta S}{\delta g^{\mu \nu}}$ for us. \\

We compute the entanglement entropy using (\ref{eq:master_formula}) in three steps : 
  \begin{enumerate}
   \item We first compute $\bra T^\mu_\mu \ket$ on the plane.
   \item Then using conformal symmetry we compute $\bra T^\mu_\mu \ket$ on the cone.
   \item Finally we insert the value of $\bra T^\mu_\mu \ket_{\text{cone}}$ 
   into (\ref{eq:master_formula}) and integrate to get the answer.
  \end{enumerate}
  \subsection*{Computation of $\bra T^\mu_\mu \ket_{\text{plane}}$}
  We work in complex plane \emph{i.e,} we change our coordinate as,
  \begin{subequations}
  \begin{align}
  z &= x_1 + i x_2 \\
  \bar{z} &= x_1 - ix_2
  \end{align}
  \end{subequations}
  Under this coordinate transformation the flat metric becomes,
  \begin{equation}
  \label{eq:metric}
  \eta_{\mu\nu} = \begin{pmatrix}
                   0 ~ & \frac{1}{2} \\
                   \frac{1}{2} ~ & 0
                  \end{pmatrix}; \quad 
  \eta^{\mu\nu} = \begin{pmatrix}
                  0 ~& 2\\
                  2 ~& 0 \\
                  \end{pmatrix}
  \end{equation}
  Therefore the trace of the energy-momentum tensor,
  \begin{equation}
  T^\mu_\mu = \eta^{\mu\nu} T_{\mu\nu} = \eta^{z\bar{z}}T_{z\bar{z}}+\eta^{\bar{z}z}T_{\bar{z}z} = 4 T_{z\bar{z}}
  \end{equation}
  as Lorentz invariance makes the energy-momentum tensor symmetric. Then (\ref{eq:ZT_connection}) becomes,
  \begin{equation}
  \label{eq:working_formula_T_zbz}
    \prd{\text{ ln}\,\mc{Z}}{\Log{a}}  = 2\, \int \td^2z \bra T_{z\bar{z}} \ket
  \end{equation}
  To compute $\bra T_{z\bar z} \ket$ on the plane using \eqref{eq:working_formula_T_zbz} we 
  compute the partition function perturbatively in $\lambda$. To do so
  we first normalize the interaction term as follows.

  Using the free theory propagator\footnote{It's evident from \eqref{eq:phi-phi} that $\phi$ doesn't
  behave as a conformal primary. Rather $\pd{\phi}$ behaves as conformal primary with scaling dimension one.
  \begin{align*}
   \langle \pd \phi(z) \pd \phi(w) \rangle = - \frac{1}{4\pi} \frac{1}{(z-w)^2}
  \end{align*}} with an IR cutoff $ R$,
  \begin{align}   \label{eq:phi-phi}
   \langle \phi(z,\bar{z}) \phi(w,\bar{w}) \rangle &=- \frac{1}{2\pi}\text{ ln}\left| \frac{z-w}{R} \right| \nonumber \\
   &= - \frac{1}{4\pi}\text{ ln} \frac{(z-w)(\bar{z}-\bar{w})}{R^2}
  \end{align}
one can notice that the interaction term can be written as,
  \begin{align}
   \label{eq:norm_vertex}
   \text{cos}(\beta \phi(z)) &= \Exp{\frac{\beta^2}{4 \pi} \Log{\frac{a}{R}}}\no{\text{cos}(\beta \phi(z))} \nonumber \\
                            &= \left(\frac{a}{R}\right)^{\frac{\beta^2}{4\pi}}\no{\text{cos}(\beta \phi(z))}
  \end{align}
where $\no {\mathcal{O}}$ represents normal ordered operator   
$\mathcal{O}$.

  The partition function upto second order in $\lambda_0$
  \begin{align} \label{eq:bare_Z}
   \mc{Z} &= \left\bra \Exp{-\int \frac{\td^2z}{a^2}\frac{\lambda_0}{2 \beta^2}\Cos{\beta \phi}} \right\ket \nonumber \\
          &= \bra 1 \ket + \frac{1}{2!} \frac{\lambda_0^2}{4 \beta^4} \left(\frac{a^2}{R^2}\right)^{\frac{\beta^2}{4\pi}}\left\bra \int \frac{\td^2z_1}{a^2}\int \frac{\td^2z_2}{a^2}\no{\Cos{\beta \phi(z_1)}}\no{\Cos{\beta \phi(z_2)}}\right\ket \nonumber \\
          &\equiv {\cal Z}_0(1 + \mc{Z}_2)
  \end{align}
  The term linear in $\lambda_0$ vanishes due to momentum conservation. ${\cal Z}_0$ is the partition 
  function of the free theory and gives the leading term $S_{EE}= {c\over 3}\ln{l\over a}$. We need 
  the contribution due to $(1+{\cal Z}_2)$.
  We set ${\cal Z}_0=1$ hereafter.
  
  The interaction term of the sine-Gordon action \eqref{eq:bare_complex_action} can be represented
  by particular vertex operator which behaves as primary operator in the theory. The scaling dimension 
  ($\Delta$) of the vertex operator can be extracted from corresponding two-point correlator. Using 
  \eqref{eq:phi-phi} it is straight forward to compute
  \begin{equation} \label{2pt-fn}
   \bra :\Cos{\beta \phi(z_1)}::\Cos{\beta \phi(z_2)}:\ket = \frac{1}{2} \left|\frac{z_1-z_2}{R}\right|^{-\beta^2/2\pi} 
\end{equation}
  Hence the operator $\Cos{\beta \phi}$ by which we deform the free conformal theory has scaling dimension,
  $\Delta = \beta^2/4 \pi$ ( conformal weight, $h=\bar h= \beta^2/4 \pi$). \\
  Inserting \eqref{2pt-fn} in \eqref{eq:bare_Z} we find 
  \begin{align}
  \label{eq:bare_Z2}
   \mc{Z}_2 &= \frac{\lambda_0^2}{128 \pi^2 \Delta^2}\left(\frac{a^2}{R^2}\right)^{\Delta} \int \frac{\td^2z_1}{a^2}\int \frac{\td^2z_2}{a^2}\bra :\Cos{\beta \phi(z_1)}::\Cos{\beta \phi(z_2)}:\ket \nonumber \\
            &=  \frac{\lambda_0^2}{128 \pi^2 \Delta^2}\left(\frac{a^2}{R^2}\right)^{\Delta}\int \frac{\td^2z}{a^2}\int \frac{\td^2w}{a^2} \frac{1}{2}\left|\frac{w}{R}\right|^{-2 \Delta} \nonumber \\
            &= \frac{\lambda_0^2}{64 \Delta^2(1-\Delta)} \left[  \left(\frac{a^2}{R^2}\right)^{\Delta-2}-\frac{R^2}{a^2}\right]
  \end{align}
  The second term inside the parentheses in \eqref{eq:bare_Z2} is badly divergent as $a\to 0$. 
  To get rid of that divergent term we add a cosmological constant term in the bare action,
  \begin{equation}
   \mc{A}_{SG} = \int \frac{\td^2z}{a^2} \left[ -\frac{\lambda_0}{\beta^2} \Cos{\beta \phi(z)} + \lambda_0(0) \right]
  \end{equation}
  The partition function becomes (upto 2nd order),
  \begin{equation}
   \mc{Z} = 1+ \frac{\lambda_0^2}{64 \Delta^2(1-\Delta)} \left[  \left(\frac{a^2}{R^2}\right)^{\Delta-2}-\frac{R^2}{a^2}\right] + \lambda_0(0)\frac{2 \pi R^2}{a^2}
  \end{equation}
  We choose
  \begin{equation}
   \lambda_0(0) = \frac{1}{2 \pi}\frac{\lambda_0^2}{64 \Delta^2(1-\Delta)}
  \end{equation}
  to cancel the divergent piece and then the partition function becomes
  \begin{equation}
   \mc{Z}_b = 1+ \frac{\lambda_0^2}{64 \Delta^2(1-\Delta)}   \left(\frac{a^2}{R^2}\right)^{\Delta-2}
  \end{equation}
  We need to renormalize the theory so that we can take the limit, $a\to 0$ smoothly and describe 
  the theory at a larger length scale $L$.  The renormalized coupling constant, $\lambda_R$ in 
  terms of the bare coupling,
  	\begin{equation}
  	\lambda_R= \lambda_0 \left(\frac{a^2}{L^2}\right)^{\frac{\Delta}{2}-1}
  	\end{equation}
  	See Appendix \ref{sec:renorm} for the detailed renormalization procedure. The renormalized
  	partition function upto the second order in $\lambda_R$ becomes,
  \begin{equation}
  \label{eq:Z_renorm}
   \mc{Z}_R = 1+ \frac{\lambda_R^2}{64 \Delta^2(1-\Delta)}   \left(\frac{L^2}{R^2}\right)^{\Delta-2}
  \end{equation}
  which is finite\footnote{Effectively we have just replaced the bare coupling $\lambda_0$ by the 
  renormalized one ($\lambda_R$) and also replaced the `lattice spacing' $a$ by a `macroscopic' or 
  `larger' length scale $L$.} at the limit $a\to 0$ and therefore one can take the continuum limit. \\  
  Using (\ref{eq:working_formula_T_zbz}) and (\ref{eq:Z_renorm}) we write the expectation value
  of the trace of renormalized energy-momentum tensor,
  \begin{align}
   2\, \int \td^2z \bra T_{z\bar{z}} \ket_{\text{pl.}} &= L\prd{\text{ ln}\,\mc{Z}_R}{L} \nonumber \\
                                                 &= \frac{\lambda_R^2}{32} \, \frac{(\Delta-2)}{\Delta^2(1-\Delta)}   \left(\frac{L^2}{R^2}\right)^{\Delta-2}
  \end{align}
  Assuming $\bra T_{z\bar{z}} \ket$ to be independent of $z$ for translational invariant system
  \bea
   \bra T_{z\bar{z}} \ket_{\text{pl.}} = \frac{\lambda_R^2}{128 \pi} \, \frac{(\Delta-2)}{\Delta^2(1-\Delta)}   \left(\frac{L^2}{R^2}\right)^{\Delta-2}\, \frac{1}{R^2}
  \eea
  
  \subsection*{Computation of $\bra T^\mu_\mu \ket_{\text{cone}}$}
  One can find the expectation value $\bra T^\mu_\mu \ket$ on a cone with angular circumference $2\pi n$ mapping 
  \be
  \label{eq:map2cone}
  \frac{w}{R}=\left(\frac{z}{R}\right)^n
  \ee
  The vacuum of the sine-Gordon theory becomes  conformally invariant when $\Delta=2$. We assume that the vacuum is still conformally invariant when the conformal weight $\Delta$ is slightly away from the marginality ($(\Delta-2)\approx 0$). Then one can use the transformation law of the sine-Gordon operator $\Cos{\beta \phi}$ under \eqref{eq:map2cone} to find the expectation value $\bra T^\mu_\mu \ket$ on the cone to $\mathcal{O}(\Delta-2)$ as 

  \begin{align}
  \bra T_{w\bar{w}}\ket_{\text{cone}}=\bra T_{\bar{w}w}\ket_{\text{cone}} &= \left(\trd{w}{z}\right)^{-\frac{\Delta}{2}} \left(\trd{\bar{w}}{\bar{z}}\right)^{-\frac{\Delta}{2}} \bra T_{z\bar{z}} (w(z),\bar{w}(\bar{z}))\ket_{\text{pl.}}\nonumber \\
                           &= n^{- \Delta} \left(\frac{w\bar{w}}{R^2}\right)^{\frac{-\Delta}{2}(1-\frac{1}{n})} \bra T_{z\bar{z}} (w(z),\bar{w}(\bar{z}))\ket_{\text{pl.}}
 \end{align}
  
  One can notice that $\bra T_{z\bar{z}} (w(z),\bar{w}(\bar{z}))\ket_{\text{pl.}}$ is linear in $(\Delta-2)$. The corrections to the above expression comes in at $\mathcal{O}\left((\Delta-2)^2\right)$.
  \subsection*{Entanglement entropy}
  To use \eqref{eq:master_formula} for computing entanglement entropy we evaluate the integral,
  \begin{align}
   \label{eq:integral}
    I(n) &=\int_{M_n} \sqrt{g} \langle T^\mu_\mu \rangle \td^2x \nonumber \\
         &= 2 \,\int_M \td w\td\bar{w} \left[n^{- \Delta} \left(\frac{w\bar{w}}{R^2}\right)^{\frac{-\Delta}{2}(1-\frac{1}{n})} (T_{z\bar{z}} (w(z),\bar{w}(\bar{z})))_{\text{pl.}}\right] \nonumber \\
      &= \frac{8\pi \, n^{2-\Delta}}{\Delta-n(\Delta-2)} \cross \bra T_{z\bar{z}}\ket_{\text{pl.}} \cross \left[ R^2 -a^2 \left(\frac{R}{a}\right)^{(\Delta-2)(n-1)}\right]  
  \end{align}
  Then entanglement entropy becomes
  \begin{align}
   \prd{\,\Delta S_{\pd A}}{\,\text{ln} a} &= \left(1-n\frac{\text{d}}{\td n}\right)_{n=1}\, I(n) \nonumber \\
                                &= 2 \pi R^2  \bra T_{z\bar{z}}\ket_{\text{pl.}} \Delta
  \end{align}
 
   The RG flow starts from a scale $a$  (UV) to $R$ (IR). $L$ is some intermediate point. 
   To avoid having too many scales  we just replace $L$ by $a$. \\

  As mentioned in the introduction, near $\Delta=2$, but with $\Delta>2$, the cosine perturbation 
  is irrelevant. The system is on the critical surface and the correlation length is infinite. Then 
  we can argue that the only other scale is the IR cutoff, which we take to be $l$ (i.e.$R=l$). Then 
  $\lambda$ is the coupling at scale $a$ and flows to zero at scale $l\gg a$.
  
  \begin{equation} \label{EE} 
   \Delta S_{\pd A} =  \frac{\lambda_R^2 (\Delta -2)}{128}  \Log{\frac{l}{a}} + \mathcal{O}\left((\Delta-2)^2\right)
   \end{equation} \\
   When $\Delta <2$, in region III, the perturbation is relevant, the correlation length $\xi$ is 
   finite   and we should replace $l\over a$ by $\xi \over a$. Similarly in region II also $\xi$ is finite. 
   In region II an expression for $\xi$ is available:  
   $\ln {\xi\over a}={4\pi \over \lambda_{int}}$  where $\lambda_{int}$ 
   is the intercept of the RG trajectory on the $y$-axis - which corresponds to the line 
   ${\beta^2\over 4\pi}-2=0$ \cite{Amit:1979ab}.
       \begin{equation} \label{EE1} 
  \Delta S_{\pd A} =  \frac{\lambda_R^2 (\Delta -2)}{128}  \Log{\frac{\xi}{a}} + \mathcal{O}\left((\Delta-2)^2\right)=\frac{\lambda_R^2 (\Delta -2)}{128}{4\pi \over \lambda_{int}}+\mathcal{O}\left((\Delta-2)^2\right)
   \end{equation}
 But as we argue below our whole analysis remains valid even for finite interval $l$ upto
 $\mathcal{O}(\Delta -2)$. For this finite interval\footnote{Note that when we compute the change 
 in EE from bulk in section \ref{sec:HEE} we consider a finite interval. Thus we should really compare 
 that holographic result with \eqref{EE} and \eqref{EE_ir}. They indeed match upto leading order in $\delta$.} case, near criticality even for $\Delta < 2$ 
 the correlation length $\xi \gg l$ and therefore the change in EE becomes
  \begin{equation} \label{EE_ir} 
  \Delta S_{\pd A} =  \frac{\lambda_R^2 (\Delta -2)}{128}  \Log{\frac{l}{a}} + \mathcal{O}\left((\Delta-2)^2\right)
  \end{equation}
which is identical to the $\Delta > 2$ case (see \eqref{EE}).  \\

 However in regions II and III clearly perturbation theory is suspect because $\lambda$ necessarily
 becomes large at the scale  of the correlation length.  \\

One can arrive at the same result \eqref{EE} by using branch point twist field method \cite{Calabrese:2004eu,Cardy:2007mb}. See appendix \ref{sec:twist}.

\section*{$\bullet$ Generalization to any finite sub-system}  
 Now we show that $\bra T_{z\bar{z}} \ket_{\text{pl.}}$ does not change to the leading order in
 $\delta$ when we map the half-line to any sub-system with arbitrary finite size using conformal map
 \begin{align}
 \label{half-line-map}
 w = f(z) = - \frac{\sin{\frac{\pi}{R}(z- l)}}{\sin{\frac{\pi}{R} z}}.
 \end{align}
 Under this transformation the action changes to
 \begin{align}
 \mathcal{A} &= \int dw \, d\bar{w} \left[ \partial_w \phi(w,\bar{w}) \,\partial_{\bar{w}} \phi(w,\bar{w}) -\frac{1}{2}\frac{\lambda_0}{\beta^2}\,\mathcal{F}(z,\bar{z})\, \text{cos}(\beta \phi(w,\bar{w}))\right] \nonumber \\
 &\equiv \mathcal{A}_0 + \mathcal{A'}_{SG}
 \end{align}
 with
 \begin{align}
 \mathcal{F}(z,\bar{z}) &= \frac{w'(z)^h \bar{w}'(\bar{z})^{\bar{h}}}{|w'(z)|^2} 
 \end{align}
 The aim is to check whether the half-line map \eqref{half-line-map} gives rise to any UV sensitive
 terms which are universal \emph{i.e,} $\Log{a}$. If it doesn't introduce any such $\Log{a}$ piece we
 can safely use this half line map. Here $z=0$ and $z=l$ are the two  UV sensitive points. We just need 
 to check if the transformation gives rise to any $\Log{a}$ contribution near those points. 
 \subsubsection*{Near $z=0$}
 The half-line map near small $z$ reduces to 
 \begin{align}
 w &= - l\,\frac{\sin{\frac{\pi}{R}(z- l)}}{\sin{\frac{\pi}{R}z}} \nonumber \\
 \label{eq:mapz0}
 &= l \left(\frac{l}{z} - 1\right) \quad \quad (for \ \  l \ll R)  
 %z &= \frac{l}{w+1}
 \end{align}
and
 \begin{align}
 \mathcal{F}(z,\bar{z}) 
 &= \left[ \left(1+\frac{w}{l}\right)\left(1+\frac{\bar{w}}{l}\right)\right]^{(\Delta-2)}
 \end{align}
 The action \eqref{eq:bare_complex_action} changes to 
 \begin{align}
 \mathcal{A} &= \int dw \, d\bar{w} \left[ \partial_w \phi(w,\bar{w}) \,\partial_{\bar{w}} \phi(w,\bar{w}) -\frac{1}{2}\frac{\lambda_0}{\beta^2 a^2}\,\left[ \left(1+\frac{w}{l}\right)\left(1+\frac{\bar{w}}{l}\right)\right]^{(\Delta-2)}\, \text{cos}(\beta \phi(w,\bar{w}))\right]
 \end{align}
 The partition function upto second order in $\lambda$
 \be
 \mc{Z} = 1+ \mc{Z}_2
 \ee
 where
 \begin{align}
 \mc{Z}_2 &= \frac{\lambda_0^2 }{256 \pi^2 \Delta^2} a^{4\delta} \int \td^2w_1 \int \td^2w_2 \left|\left(1+\frac{w_1}{l}\right)\left(1+\frac{w_2}{l}\right)\right|^{4 \delta} \frac{1}{|w_1 - w_2|^{2 \Delta}}\nonumber \\
 \end{align}
 Near marginality i.e. $\Delta=2$ or $\delta = 0$ the leading term in the above expression becomes
 \begin{align}
  \mc{Z}_2 &\approx \frac{\lambda_0^2 }{256 \pi^2 \Delta^2} a^{4\delta} \int \td^2w_1 \int \td^2w_2 \, \{1  + 2 \delta [\Log{1+\frac{w_1}{l}}+ \Log{1+\frac{\bar w_1}{l}}+\Log{1+\frac{w_2}{l}} \nonumber\\ 
  \label{eq:z2lin}
  &\hspace{6cm} + \Log{1+\frac{\bar w_2}{l}} ]  \}  \, \frac{1}{|w_1 - w_2|^{2 \Delta}}
 \end{align}
 The $\mc{O}(\delta^0)$ term in the integral of the above expression is what we computed in \eqref{eq:bare_Z2}.
 The terms inside the square brackets  (which are $\mc{O}(\delta)$) have appeared due to the half line map. 
 We need to show that these terms don't give rise to any UV sensitive terms which are universal up to $\mc{O}(\delta)$.\\
 \subsubsection*{Near $ z=l $}
 The half-line map near $z=l$ reduces to
 \begin{align}
 w &\approx -l\, \frac{\frac{\pi}{R}(z- l)}{\sin{\frac{\pi}{R}z}} \nonumber \\
 &= l\left( \frac{l}{z} - 1 \right)\quad \quad (for \ \  l \ll R)  
 %z &= \frac{l}{w+1}
 \end{align}
 As the map remains same as \eqref{eq:mapz0} the entire analysis in the above section holds true in 
 this case too. Therefore we essentially need to show that the integral
 \begin{equation}
 \label{eq:I}
 I= 2\delta \int \td^2w_1 \int \td^2w_2 \left[\Log{1+\frac{w_1}{l}}+ \Log{1+\frac{\bar w_1}{l}} \right]\, \frac{1}{|w_1 - w_2|^{2 \Delta}}
 \end{equation}
 in \eqref{eq:z2lin} does not result in any log-divergence to $\mathcal{O}(\delta) $. Then we can claim 
 that to $\mathcal{O}(\delta) $ our answer for $\bra T_{z\bar{z}} \ket_{\text{pl.}}$ still holds true for 
 any finite subsystem of size $l$.
 
 Note  that to $\mathcal{O}(\delta) $ the contribution from \eqref{eq:I} is UV finite. Because, 
 at UV region,  ($i.e, \ w_1 \approx w_2 $) the divergence comes only from the factor
 $ \frac{1}{|w_1 - w_2|^{2 \Delta}} $. But the contribution is  $\mathcal{O}(\delta^2) $ 
 since there is already an extra $\delta$ sitting outside the integral. Further more at $ w_1 \to 0 $ 
 the integral is finite and hence there is no log divergent piece.
 
 Thus  our answer for entanglement entropy holds true for any sub-system with size $l$ as the 
 value of $\bra T_{z\bar{z}} \ket_{\text{pl.}}$ does not change under the conformal map to leading order in $\delta$.

\section{\label{sec:HEE}Entanglement entropy from holography}
 
In this section we compute the holographic entanglement entropy (HEE) for a single interval 
in two dimensional theory by Ryu-Takayanagi prescription. The 2D theory is not conformal but 
deformed by a primary  operator with conformal dimension $\Delta$. Our goal is to check whether 
this change in HEE due to the deformation matches \footnote{It is worth mentioning that we are not claiming
$AdS_3$ with a massive scalar is dual to sine-Gordon theory in 1+1 dimensions. Moreover the way we compute 
EE from bulk and the boundary theory, they are both in weakly coupled regime. As we will see in this section 
the results match only upto leading order in $\lambda_0$ (possibly due to some `universality') and there is 
no reason for them to match at higher orders. We come back to this point in section \ref{sec:conclusions}} 
our field theory result at least leading order in the coupling $\lambda_0.$ 

  According to the holographic dictionary, insertion of a primary operator of scaling dimension $\Delta$ 
  in the boundary theory
 can be realized by including a free massive scalar field in the bulk action of mass, $m$ such that
 $m^2= \Delta(\Delta-2)$. The scalar field back-reacts and changes the metric. Under this metric 
 perturbation the holographic entanglement entropy also changes. 
 
   \vspace{1cm}
  \begin{center}
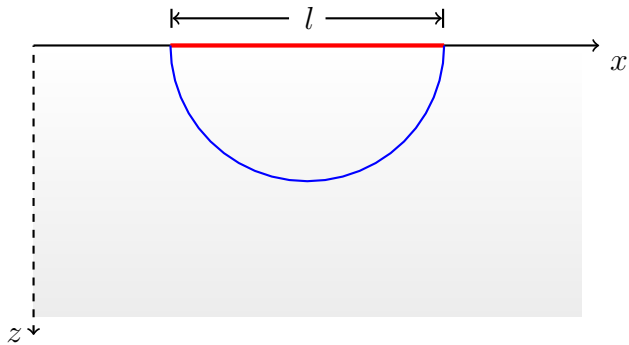

    \begin{tikzpicture}[scale =1.2]

\fill[top color=white,bottom color=gray!15] (-3,0)--(3,0)--(3,-3)--(-3,-3);

\draw [->, thick] (-3,0) -- (3.2,0);
\draw [<-, dashed, thick] (-3,-3.2) --(-3,0);

\node [left] at (-3,-3.2) {$z$};
\node [below right] at (3.2,0) {$x$};
\draw [ultra thick,red,fill=yellow] (-1.5,0) --(1.5,0);
\draw [|<-, thick] (-1.5,0.3) --(-0.2,0.3);
\draw [->|, thick] (.2,0.3) --(1.5,0.3);
\node [left] at (.2,0.3) {$l$};

   \draw [blue,thick,domain=180:360] plot ({1.5*cos(\x)}, {1.5*sin(\x)});

\end{tikzpicture}

\captionof{figure}{\emph{According to Ryu-Takayanagi prescription EE of a segment in CFT$_2$ is given by 
the corresponding geodesic length in AdS$_3$. For slightly perturbed CFT at leading order the EE is given by 
the same semi-circle but its length is changed due to change in bulk  metric.}} \label{fig:bulk_EE}

  \end{center}
 
 \vspace{0.5cm}

 Upto first order change in metric
 \begin{equation}
  g_{\mu\nu} = \bar g_{\mu\nu} + h_{\mu\nu} 
 \end{equation}
 the holographic entanglement entropy changes as
 \begin{equation}
 \label{eq:master_eq}
  \Delta S_{\pd A} = \frac{1}{8 G_N} \int_A \td \xi^{(d-2)} \sqrt{G^{(0)}}G^{(0)ij}G^{(1)}_{ij}
 \end{equation}
 where $\xi$'s are coordinates on the co-dimension two extremal surface, $G_N$ is the $d$-dimensional 
 Newton's constant and $G^{(0)} = \text{det} G^{(0)}_{ij}$ and the induced metric and its variation 
 are given by
 \begin{align}
  G^{(0)}_{ij} &= \prd{x^\mu}{\xi^i} \prd{x^\nu}{\xi^j} \bar g_{\mu\nu} \\
  G^{(1)}_{ij} &= \prd{x^\mu}{\xi^i} \prd{x^\nu}{\xi^j} h_{\mu\nu}
 \end{align}
 Under the metric perturbation the extremal surface also changes
 \begin{equation}
  z(\xi) = z_0(\xi)+z_1(\xi)+\cdots
 \end{equation}
 To the first order we consider only $z(\xi) = z_0(\xi)$.  $z_1(\xi)$ contributes from second 
 order because by definition of extremality the first order change to the length is zero. The 
 contribution due to $\bar g_{\mu\nu}$
 is the original Ryu-Takayanagi calculation that gives $S_{EE}= {c\over 3}\ln {l\over a}$.
 
 For our case $(d=3)$, the background metric, $\bar g_{\mu\nu}$ in Poincare patch is
 given by,
 \begin{align}
  ds^2_{AdS_3} &=\bar g_{\mu\nu} dx^{\mu} dx^{\nu} \nonumber \\
               &= \frac{dz^2+dx_1^2+dx_2^2}{z^2}
 \end{align}
 We fix AdS radius $R_{AdS}=1$. We also take the fluctuation to be
\begin{equation}
 h_{\mu\nu} = \frac{1}{z^2} \chi_{\mu\nu}
\end{equation}

 The co-dimension two surface is one dimensional. We take $\xi^1$ to be $x^1$ 
 and denote it as $x$. The profile of the extremal surface is given by
 \begin{equation}
  z_0(x) = \frac{1}{2} \sqrt{l^2-4 x^2}
 \end{equation}

 Then (\ref{eq:master_eq}) becomes
 \begin{align}
 \label{eq:hee}
  \Delta S_{\pd A} &= \frac{1}{8 G_N} \int_A \td x \, \sqrt{G^{(0)}}\,G^{(0)xx}\,G^{(1)}_{xx} \nonumber \\
                   &= \frac{1}{4 G_N l} \int_{-l/2}^{l/2} \td x \left[z_0^2 \, G^{(1)}_{xx}\right]
 \end{align}
 using
 \begin{align}
 G^{(0)}_{xx} &= \left(\prd{z}{x}\right)^2 \bar g_{zz}+\bar g_{xx} \nonumber \\
 &= \frac{l^2}{4}\frac{1}{z_0^4}
 \end{align}

 To obtain $\Delta S_{\pd A}$ we need to compute $G^{(1)}_{xx}$. We find $G^{(1)}_{xx}$ solving 
 linearized Einstein equation for $h_{\mu\nu}$ considering the back-reaction due to the massive scalar field.
 \subsubsection*{Linearized Einstein equation :}
 Consider the linearized Einstein equation due to the back-reaction of a massive scalar field $\Phi$ on 
 \AdS background.
 \begin{align}
 \label{eq:lin_cov_ein_eq}
  -\frac{1}{2} \bar \nabla^2 \tilde h_{\mu\nu}+ \frac{1}{2} \left(\bar \nabla_\mu \bar \nabla^\lambda \tilde h_{\lambda\nu}+\bar \nabla_\nu \bar \nabla^\lambda \tilde h_{\lambda\mu}\right)-\frac{1}{2}\bar g_{\mu\nu} \bar \nabla^\alpha \bar \nabla^\beta \tilde h_{\alpha\beta}-(\tilde h_{\mu\nu}- \bar g_{\mu\nu} \tilde h) = 8\pi G_N \, T^{(1)}_{\mu\nu}
 \end{align}
 where 
 \bea
  \tilde h_{\mu\nu} = h_{\mu\nu}-\frac{1}{2}\bar g_{\mu\nu} h 
   \eea
\bea
 \tilde h = \bar g^{\mu\nu}\tilde h_{\mu\nu}; \qquad  h = \bar g^{\mu\nu} h_{\mu\nu} 
 \eea
 The stress-energy tensor for the massive scalar field is given by,
 \begin{equation}
 \label{eq:stress_tensor_scalar}
  T^{(1)}_{\mu\nu} = \pd_{\mu}\Phi \, \pd_{\nu}\Phi -\tfrac{1}{2} \bar g_{\mu\nu} 
  \left(\bar g^{\alpha\beta} \, \pd_{\alpha}\Phi \, \pd_{\beta}\Phi+m^2 \Phi^2 \right)
 \end{equation}
 The mass of the scalar field is related to the scaling dimension $\Delta$ of the boundary operator 
 as
 $\Delta(\Delta-2)=m^2$.
 \subsubsection*{Gauge choice:}
 Equation (\ref{eq:lin_cov_ein_eq}) is invariant under diffeomorphism generated by the vector field,
 $\xi$
 \begin{equation}
  h_{\mu\nu} \rightarrow h_{\mu\nu}+\bar \nabla_{\mu} \xi_{\nu}+\bar \nabla_{\nu} \xi_{\mu}
 \end{equation}
 Clearly we need $D=3$ independent constraint equations to fix the gauge. We choose
 \begin{equation}
  h_{z\mu} = 0
 \end{equation}
 in the Poincare patch.
 \subsubsection*{Constraints on the bulk field:}
 The gauge/gravity correspondence states that turning on a bulk field which behaves as 
 $z^{2-\Delta} \lambda_{b}(x)$ near the boundary $z \to 0$ is dual to  a
 source term $ \int \lambda_b(x) \mathcal{O}(x)$ in the CFT action, where $\mathcal{O}(x)$ is a 
 CFT operator. In our case, $\lambda_b(x)$ does not depend
 on $x$. This imposes the condition that the bulk field, $\Phi(z,x)$ is also independent of $x$. 
 Since the stress-energy tensor of the bulk field $\Phi$ is sourcing the equation (\ref{eq:lin_cov_ein_eq})
 $h_{\mu\nu}$ also does not depend on the boundary coordinates.
 \subsubsection*{Linearized equation in Poincare patch:}
 Imposing the gauge choice as well as the constraints on the bulk fields in Poincare patch the equation
 (\ref{eq:lin_cov_ein_eq}) takes the form
 \bea
 \label{eq:lin_ein_eq_pncr_1}
  \frac{1}{z}[\chi_{xx}'(z)+\chi_{yy
   }'(z)] = - 16 \pi G_N  \, T^{(1)}_{zz} \\ \nonumber \\
  \label{eq:lin_ein_eq_pncr_2}
 \chi_{yy}''(z)-\frac
   {\chi_{yy}'(z)}{z}
   =  16 \pi G_N \, T^{(1)}_{xx} \\ \nonumber \\
  \label{eq:lin_ein_eq_pncr_3}
  \chi_{xy}''(z)-\frac
   {\chi_{xy}'(z)}{z} = - 16 \pi G_N  \, T^{(1)}_{xy} \\ \nonumber \\
  \label{eq:lin_ein_eq_pncr_4}
   \chi_{xx}''(z)-\frac
   {\chi_{xx}'(z)}{z}= 16 \pi G_N \, T^{(1)}_{yy} \\ \nonumber \\
  \label{eq:lin_ein_eq_pncr_5}
  16 \pi G_N \, T^{(1)}_{zx} = 16 \pi G_N \, T^{(1)}_{zy} = 0  
 \eea
 \subsubsection*{Bulk equation of motion of a massive scalar field:}
 To solve \eqref{eq:lin_ein_eq_pncr_1}-\eqref{eq:lin_ein_eq_pncr_5} we need to know the functional 
 form of $T^{(1)}_{\mu\nu}$.
 We solve the bulk equation of motion of the scalar field and pick the non-normalizable mode
 which we substitute in (\ref{eq:stress_tensor_scalar}) to get the RHS of 
 \eqref{eq:lin_ein_eq_pncr_1}-\eqref{eq:lin_ein_eq_pncr_5}.
 The equation of motion is given by
 \begin{equation}
  \frac{1}{\sqrt{\bar g}} \,\pd_\mu \left(\sqrt{\bar g}\, \bar g^{\mu\nu} \,\pd_\nu \Phi \right) - m^2 \Phi =0
 \end{equation}
 Imposing the constraint on $\Phi$ the above equation takes the form
 \begin{equation}
 \label{eq:scalar_eq_motion}
  z^2 \frac{\td^2 \Phi}{\td z^2} - z \frac{\td \Phi}{\td z}-m^2 \Phi =0
 \end{equation}
 The general solution to the above equation is given by,
 \begin{equation}
  \Phi(z) = c_1 z^{1+\sqrt{1+m^2}} + c_2 z^{1-\sqrt{1+m^2}}
 \end{equation}
 Writing $\Delta = 1 + \sqrt{1+m^2}$, the non-normalizable mode reduces to
 \begin{equation}
 \label{eq:non-norm_mode}
  \Phi(z)_{nn} = \frac{\lambda_b}{4\pi \Delta} \left(\frac{z}{a}\right)^{2-\Delta}
 \end{equation}
 where we have used the boundary condition, $\Phi(z=a) = \lambda_b/4\pi\Delta$. Here $\lambda_b$ 
 is the coupling constant of the sine-Gordon interaction in the boundary up to a normalization factor
 (see Appendix \ref{sec:norm}).

  Note that for $\Delta <2$ (relevant perturbation) the the non-normalizable mode $\Phi(z)_{nn}$ 
  blows up in the interior and also for $\Delta > 2$ (irrelevant perturbation) the non-normalizable 
  mode $\Phi(z)_{nn} \to \infty$ if one takes $z \to 0$. See equation \eqref{eq:non-norm_mode}. Therefore
  for both situations the theory has to be modified. But computing EE in the three dimensional bulk amounts
  to computing geodesic length which gets most of its contribution from near boundary ($z=a$) region. 
  Therefore as long as one is interested in EE the divergence in the deep interior (IR divergence) is not 
  important. On the other hand since we cut off the geometry and place our boundary theory at $z = a$ our 
  result is insensitive to the UV divergence near $z=0$. Thus the change in EE we compute is valid both for 
  relevant and irrelevant perturbations. 
  
 \subsubsection*{Stress-energy tensor:}
 Using (\ref{eq:non-norm_mode}) we find
 \begin{align}
  T^{(1)}_{zz} &= \pd_z\Phi \,\pd_z\Phi - \frac{1}{2 z^2}\left( \bar g^{\alpha\beta} \, \pd_{\alpha}\Phi \, \pd_{\beta}\Phi+m^2 \,\Phi^2 \right) \nonumber \\
               &= -\left(\frac{\lambda_b}{4\pi\Delta~a}\right)^2 (\Delta-2) \left(\frac{z}{a}\right)^{2(1-\Delta)}
 \end{align}
 and
 \begin{align}
  T^{(1)}_{xx}=T^{(1)}_{yy} &= - \frac{1}{2 z^2}\left( \bar g^{\alpha\beta} \,\pd_{\alpha}\Phi \, \pd_{\beta}\Phi+m^2 \, \Phi^2 \right) \nonumber \\
               &= -\left(\frac{\lambda_b}{4\pi\Delta~a}\right)^2 (\Delta-2) (\Delta-1)\left(\frac{z}{a}\right)^{2(1-\Delta)}
 \end{align}
 \begin{equation}
  T^{(1)}_{zx} = T^{(1)}_{zy}=T^{(1)}_{xy} = 0.
 \end{equation}
 \subsubsection*{Differential equations to solve:}
 Clearly \eqref{eq:lin_ein_eq_pncr_5} are trivially satisfied.
 The remaining equations become\
 \bea
  \frac{1}{2}
   \left(\chi_{xx}''(z)-\frac
   {\chi_{xx}'(z)}{z}\right) &=& - \, 8\,\pi\, G_N \left(\frac{\lambda_b}{4\pi\Delta~a}\right)^2 (\Delta-2) (\Delta-1)\left(\frac{z}{a}\right)^{2(1-\Delta)} \nonumber \\ \nonumber \\ 
  \frac{1}{2}
   \left(\chi_{yy}''(z)-\frac
   {\chi_{yy}'(z)}{z}\right)  &=& -\, 8\, \pi \, G_N  \left(\frac{\lambda_b}{4\pi\Delta~a}\right)^2 (\Delta-2) (\Delta-1)\left(\frac{z}{a}\right)^{2(1-\Delta)} \nonumber \\ \nonumber  \\
   \frac{\chi_{xx}'(z)+\chi_{yy}'(z)}{ z} &=& 16 \, \pi \, G_N \left(\frac{\lambda_b}{4\pi\Delta~a}\right)^2 (\Delta-2) \left(\frac{z}{a}\right)^{2(1-\Delta)} \nonumber \\ \nonumber  \\  
 \chi_{xy}''(z)&-&\frac{\chi_{xy}'(z)}{z} = 0 \nonumber  
 \eea

 The solution for $\chi_{xx}(z)$ is given by
 \begin{equation} \label{metric_soln}
  \chi_{xx}(z)=-(4\pi G_N )\frac{\lambda_b^2}{16\pi^2\Delta^2} \left(\frac{z^2}{a^2}\right)^{2-\Delta} +C_1 z^2  + C_2
 \end{equation}
The constant term\footnote{Note that we perturb $\bar{g}_{\mu \nu}$ by $\frac{1}{z^2} \chi_{\mu \nu}$. 
Near the boundary when perturbation is close to marginality : 
$\frac{1}{z^2} \chi_{x x} \sim  \frac{\#}{z^2} + C_1  + \frac{C_2}{z^2}$. Clearly the first and the
third terms in \eqref{metric_soln} modify the fall-off behavior near the boundary. This is expected 
since we are working with non-nomalizable mode which changes the asymptotic geometry. This modification 
in boundary condition is dual to deformation of the boundary theory.} would give us change in EE, 
$\Delta S_{\pd A} = const.$ which is not of interest to us and therefore we drop $C_2$. We are working 
in a regime where the perturbation is very close to be marginal \emph{i.e,} $\Delta = 2 + 2 \delta$. 
Since $\delta$ is very close to zero near the boundary ($z=0$) the first term is always dominant 
compared to the second term in Eqn. \eqref{metric_soln}. Thus 
$ \chi_{xx}(z)$ involves only the particular solution to the corresponding differential equation
\begin{equation} \label{metric_soln}
  \chi_{xx}(z)=-(4\pi G_N )\frac{\lambda_b^2}{16\pi^2\Delta^2} \left(\frac{z^2}{a^2}\right)^{2-\Delta}
 \end{equation}

 For $\Delta > 2$ the perturbation is irrelevant\footnote{According the ``holographic RG'' $z=0$ 
 corresponds to the UV theory and $z \to \infty$ corresponds to IR. For $\Delta = 2 + 2 \delta$, 
 the term (which is the field theory coupling profile)  decays as $z^{-4 \delta}$ near $z \to \infty$.
 Therefore the perturbation becomes irrelevant for IR physics. By the same argument, 
 for $\Delta = 2 - 2 \delta$, the coupling grows as $z^{4 \delta}$ near $z \to \infty$. 
 Making the perturbation more important near low-energy scale.} whereas the  $\Delta < 2$ indicates 
 relevant perturbation.  

 \subsubsection*{Change in entanglement entropy:}
 As $h_{zz}=0$ by our gauge choice, then $G^{(1)}_{xx}= \frac{1}{z_0^2}\chi_{xx}$. 
 Then \eqref{eq:hee} becomes
 \begin{align}
  \Delta S_{\pd A} &= \frac{1}{4 G_N l} \int_{-l/2}^{l/2} \td x  ~\chi_{xx} \nonumber \\
                   &= -\frac{\pi \lambda_b^2}{16\pi^2\Delta^2~l} \int_{-l/2}^{l/2} \td x~  \left(\frac{l^2-4 x^2}{4 a^2}\right)^{2-\Delta}\\
                   &= -\frac{\lambda_b^2}{8\pi\Delta^2} \left(\frac{l^2}{a^2}\right)^{2-\Delta} \cross \int_0^{1/2}\td y~ \left( \frac{1-4 y^2}{4}\right)^{2-\Delta}
 \end{align}

It is worthwhile to note that $G_N~$s get canceled and this makes $\Delta S_{\pd A}$ independent 
of $G_N$. This is same as $\Delta S_{\pd A}$ being independent of central charge $c$ (we have taken $R_{AdS} =1$). 

 Expanding the above result near $\Delta=2$ we find
 \bea
  \Delta S_{\pd A} =\lambda_b^2\left[ (const. ~terms)  +\frac{(\Delta-2)}{32 \pi} \Log{\frac{l}{a}} +\mathcal{O}((\Delta-2)^2) \right]
\eea

From AdS/CFT dictionary using the normalization (see Appendix \ref{sec:norm})
\bea
 \lambda_R = \lambda_b \cross \frac{2}{\sqrt{\pi}}(\Delta-1)
\eea
we get
\bea \label{HEE}
 \Delta S_{\pd A}=  \lambda_R^2\frac{(\Delta-2)}{128}\Log{\frac{l}{a}}+\mathcal{O}((\Delta-2)^2)
\eea

This expression \eqref{HEE} for EE from holography reproduces the 
result \eqref{EE} and \eqref{EE_ir} we obtained in the previous section using field theoretic technique near $\Delta = 2$. 

\section{\label{sec:conclusions}Summary and Conclusions}

In this paper we have calculated the change in EE along an RG flow near the trivial fixed point of 
the 1+1 dimensional sine-Gordon theory.  It has been calculated both in the boundary and in the 
bulk and the answers agree. 
As mentioned in the introduction one can calculate the change in central charge function along the RG flow
using the ERG \cite{Oak:2017trw}. This gives 
\[
\Delta c = c(\lambda)-c(0)={3\lambda_R^2(\Delta-2)\over 128}
\]
If we carry over the expression 
\[
\Delta S = {\Delta c \over 3} \, \ln\,\left({l\over a}\right)
\]
which holds for a CFT, 
we see that the answer agrees exactly with the calculation in this paper. Here it is worth emphasizing we are \emph{not} claiming that EE of a perturbed CFT, in general, is simply given $\frac{c(r)}{3} \,ln \left(\frac{l}{a}\right)$ where $r$ is some RG parameter. It is just a leading order effect and its form will be rather complicated at higher orders. Further, in a more computational level,  it is clear from \eqref{eq:ee_twist} that the $ln \left(\frac{l}{a}\right)$ term appears due to series expansion\footnote{In fact this well known in the literature that in CFTs perturbed by non marginal terms, the log terms arise by an expansion of power terms (see e.g, \cite{Amit:1979ab,Das:1986cz}).} about $\delta=0$.

Our field theory computation is for a single boson ($c=1$) coupled to  sine-Gordon potential. 
We have perturbatively computed the change in EE of an interval due to the interaction term.
Whereas from the bulk we have calculated the same quantity using holographic dictionary. 
It is very interesting to note when we talk about bulk geometry we are implicitly assuming 
$\,c\,$ to be very large \cite{Brown:1986nw}. Therefore the  matching of $\Delta S_{\pd A}$ might
look mysterious. The possible explanation is as follows\footnote{We thank Nemani Suryanarayana 
for pointing this out.}. The agreement of results from the bulk and the boundary side holds true only 
upto leading order in the coupling. As mentioned above, in the bulk computation the only dimensionful
parameter $G_N$ (since $R_{AdS}=1$) cancels out and consequently the change in EE (which is also equivalent 
to change in central charge, $\Delta c$) becomes independent of the value of $\,c\,$. This  `universality'
of the leading order correction makes it possible to compare and match the results from both sides. Again at higher 
orders, presumably the central charge $\,c\,$ will be important and thus the results will differ.  

The final expression for $\Delta S$ is proportional to $\Delta-2$. This is expected since one expects 
the central charge and entanglement entropy to decrease along an RG flow, because degrees of freedom 
are being integrated out. Thus if the coupling $\lambda$ is relevant (i.e. $\Delta-2<0$) then one expects
the central charge to be larger at $\lambda=0$ than at $\lambda>0$. If $\Delta-2>0$ the flow is in the opposite
direction and the central charge is larger when $\lambda>0$.

We have also seen in the boundary calculation, that the case $\Delta-2<0$, when the Cosine perturbation
grows larger in the IR, perturbation theory is harder to justify. In the bulk this effect shows up as a 
scalar field that becomes larger in the AdS interior. In this case one cannot ignore the non linear terms 
in the scalar field EOM.  In the boundary we replaced
$\ln ({l\over a})$ by $\ln ({\xi\over a})$ on intuitive grounds. We do not have a similar argument 
for the bulk. This case requires a more exact treatment using the full RG in the boundary and correspondingly
the full non linear EOM in the bulk.

 In conclusion, the computation described in this paper extends the
AdS/CFT correspondence in EE to non conformal backgrounds - but remaining close to conformality.
It is a challenge to extend this calculation to a finite distance along the RG flow. 

%%%%%%%%%%%%%%%%%%%%%%%%%%%%%%%%%%%%%%%%%%%%%%%%%
\acknowledgments
{We thank Nemani Suryanarayana for sevaral useful discussions on various aspects of the problem.
PB and AB thank  Sk Jahanur Hoque, Nirmalya Kajuri and Alok Laddha for fruitful discussions.}

%%%%%%%%%%%%%%%%%%%%%%%%%%%%%%%%%%%%%%%%%%%%%%%%%

%%%%%%%%%%%%%%%%%%%%%%%%%%%%%%%%%%%%%%%%%%%%%%%%%
\appendix
%%%%%%%%%%%%%%%%%%%%%%%%%%%%%%%%%%%%%%%%%%%%%%%%%

%%%%%%%%%%%%%%%%%%%%%%%%%%%%%%%%%%%%%%%%%%%%%%%%%
\section{\label{sec:renorm}Renormalization of coupling $\lambda$}

Here we renormalize the coupling $\lambda$ such that we can smoothly take the `continuum limit' 
$a \to 0$. We perform it in  two steps following \cite{Sathiapalan:2009ft}.  

\subsubsection*{Intermediate case}
  Here we define an intermediate coupling, $\lambda_I$ that absorbs the $a$-dependence that arises from
  normal ordering of vertex operator (\ref{eq:norm_vertex}),
  \begin{equation}
   \int \frac{\td^2 z}{a^2} \frac{\lambda_0}{4 \pi \Delta} \Cos{\beta \phi(z)} := 
   \int \frac{\td^2 z}{L^2} \left( \frac{L^2}{R^2}\right)^{\frac{\Delta}{2}}\frac{\lambda_I}{4 \pi \Delta} :\Cos{\beta \phi(z)}:
  \end{equation}
  Thus
  \begin{equation}
   \lambda_0 = \lambda_I  \left( \frac{L^2}{a^2}\right)^{\frac{\Delta}{2}-1}
  \end{equation}
  and
  \begin{equation}
   \lambda_0(0) = \lambda_I(0) \frac{a^2}{L^2}.
  \end{equation}
  Therefore the interaction term in the action becomes,
  \begin{equation}
   \mc{A}_{SG} = \int \frac{\td^2z}{L^2} \left[ -\frac{\lambda_I}{4 \pi \Delta} :\Cos{\beta \phi(z)}: \left( \frac{L^2}{R^2}\right)^{\frac{\Delta}{2}}+ \lambda_I(0) \right]
  \end{equation}
  \subsubsection*{Full renormalization}
  So all the $a$-dependences are collected in $\lambda_I$ and $\lambda_I(0)$. 
  We define a fully renormalized coupling $\lambda_R$ so that we can write the action without any 
  $a$-dependence.
  \begin{subequations}
   \begin{align}
    \lambda_I &:= \lambda_R + \delta  \lambda_R \\
    \lambda_I(0) &:= \lambda_R(0) +  \delta  \lambda_R(0)
   \end{align}
  \end{subequations}
  Thus 
  \begin{equation}
   \mc{A}_{SG} = \int \frac{\td^2z}{L^2} \left[ -\frac{\lambda_R + \delta  \lambda_R }{4 \pi \Delta} :\Cos{\beta \phi(z)}: \left( \frac{L^2}{R^2}\right)^{\frac{\Delta}{2}}+ \lambda_R(0) +  \delta  \lambda_R(0)\right]
  \end{equation}
  Now we choose the counterterms $\delta  \lambda_R $ and $\delta  \lambda_R(0) $ such that 
  the $a$-dependence is removed in order by order.
  We see that $\delta  \lambda_R $ comes into play at $\mc{O}(\lambda_R^3)$.
  We set $\lambda_R(0)=0$ and choose $\delta  \lambda_R(0)$ to cancel divergence,
  \begin{equation}
   \frac{\delta  \lambda_R(0)}{L^2} = \frac{\lambda_0(0)}{a^2}
  \end{equation}
  %

%%%%%%%%%%%%%%%%%%%%%%%%%%%%%%%%%%%%%%%%%%%%%%%%%%
\subsection*{\label{sec:beta}$\beta$-function}
  One can check whether the renormalized partition function is scale invariant by computing
  $\beta$-function. To the second order,
  \begin{equation}
   \lambda_R = \lambda_I = \lambda_0 \left(\frac{a^2}{L^2}\right)^{\frac{\beta^2}{8\pi}-1}
  \end{equation}
  Therefore,
  \begin{equation}
   \beta_{\lambda_R} = L \prd{\lambda_R}{L} = \left(2 - \frac{\beta^2}{4\pi}\right)\lambda_R.
  \end{equation}
  Now computing,
  \begin{equation}
   \prd{\mc{Z}_R}{\lambda_R} = \pi^2 \frac{\lambda_R}{(\beta^2)^2}\left(\frac{L^2}{R^2}\right)^{\frac{\beta^2}{4\pi}-2}\frac{1}{2-\beta^2/2\pi}
  \end{equation}
  and
  \begin{equation}
  \label{eq:L_del_Z_del_L}
   L\prd{\mc{Z}_R}{L} = \pi^2 \left( \frac{\beta^2}{4\pi}-2\right) \frac{\lambda_R^2}{(\beta^2)^2}\left(\frac{L^2}{R^2}\right)^{\frac{\beta^2}{4\pi}-2}\frac{1}{2-\beta^2/2\pi}
  \end{equation}
  we check that
  \begin{equation}
   L\trd{\mc{Z}_R}{L} = L\prd{\mc{Z}_R}{L} + \beta_{\lambda_R} \prd{\mc{Z}_R}{\lambda_R} =0
  \end{equation}
  \emph{i.e,} the renormalized partition function is scale independent upto second order in coupling.
  %

%%%%%%%%%%%%%%%%%%%%%%%%%%%%%%%%%%%%%%%%%%%%%%%%%%%%
%

\section{\label{sec:twist}Branch point twist fields method}
Here we show that our main result \eqref{EE} can also be derived using branch point twist fields method\footnote{The authors would like to thank the referee for his/her detailed suggestions that prompted the addition of this section.} \cite{Cardy:2007mb,CastroAlvaredo:2011zs,CastroAlvaredo:2008kh,CastroAlvaredo:2008pf,CastroAlvaredo:2009ub,Doyon:2008vu,Bianchini:2015uea,Bianchini:2016mra,Calabrese:2004eu}.

Renyi entropy is defined by,
\be 
S_n(l) = -\frac{\partial}{\partial n} \text{ Tr }\rho^n
\ee 
One gets entanglement entropy from Renyi entropy by taking $n\to 1$ limit. Using twist fields ($\mathcal{T}, \widetilde{\mathcal{T}}$) defined on replica sheet one can compute the Renyi entropy as
\be
\text{ Tr }\rho^n = a^{4\Delta_n} \langle \, \mathcal{T}(0)\widetilde{\mathcal{T}}(l) \, \rangle
\ee
where $\Delta_n$ is the conformal weight of the twist field \cite{Calabrese:2004eu},
\be
\Delta_n = \frac{c}{24}\left(n-\frac{1}{n}\right)
\ee
where $c$ is the central charge of the CFT. The OPE for the twist fields is given by
\bea
\mathcal{T}(0)\widetilde{\mathcal{T}}(l) \sim l^{-4\Delta_n} \mathbf{1}+\sum_{i} C_i \, l^{2\Delta_i -4\Delta_n} \,\mathcal{O}_i
\eea
where $\mathcal{O}_i$s are the local fields of the replica CFT with the conformal dimension $\Delta_i$ and $C_i$ are the three point coupling which depends on the operator and the theory under consideration. At criticality $\langle \mathcal{O}_i \rangle=0$.
Therefore,
\bea
S_n(l) = -\frac{\pd}{\pd n} \left(\frac{a}{l}\right)^{4\Delta_n}
\eea
and taking $n\to 1$ limit we find
\bea
S &=& \lim_{n\to 1} S_n(l) \nonumber \\
&=& \lim_{n\to 1} \frac{c}{6}\text{ ln}\left(\frac{l}{a}\right) \left(1+\frac{1}{n^2}\right)+ \mathcal{O}(n-1) \nonumber \\
&=& \frac{c}{3}\text{ ln}\left(\frac{l}{a}\right)
\eea
But away from criticality $\langle \mathcal{O}_i \rangle \neq 0$ and $C_i$'s are no longer constant. When a CFT is perturbed by a local operator with conformal dimension $\Delta$,
\bea
\mathcal{A} = \mathcal{A}_{CFT} +\lambda \int \td^2z \, \mathcal{O}(z)
\eea
the OPE coefficients are given by the following expansion \cite{Bianchini:2015uea} 
\bea
C_i \left(\frac{l}{a}\right) = C_i \left(1+ C_i^1(n) \left(\frac{l}{a}\right)^{2-\Delta}+ C^2_i \left(\frac{l}{a}\right)^{2(2-\Delta)}+\cdots\right)
\eea
where $C_i^1, C_i^2,\dots$ are constant with respect to $(l/a)$.

Leading order correction comes from the OPE with the identity operator.
\bea
\mathcal{T}(0)\widetilde{\mathcal{T}}(l) \sim l^{-4\Delta_n}\left(1+ C_1^1(n) \left(\frac{l}{a}\right)^{2-\Delta}+\cdots\right) \mathbf{1}
\eea
This implies 
\bea
a^{4\Delta_n} \langle \, \mathcal{T}(0)\widetilde{\mathcal{T}}(l) \, \rangle = \left(\frac{a}{l}\right)^{4\Delta_n}\left(1+ C_1^1(n) \left(\frac{l}{a}\right)^{2-\Delta}\right)
\eea 
Therefore
\bea
S_n(l) = -\frac{\pd}{\pd n} \left(\frac{a}{l}\right)^{4\Delta_n}-\frac{\pd}{\pd n} C_1^1(n) \left(\frac{l}{a}\right)^{2-\Delta}
\eea
Near marginality entanglement entropy becomes
\bea
\label{eq:ee_twist}
S = \frac{c}{3}\text{ ln}\left(\frac{l}{a}\right) + A + B(\Delta-2)\text{ ln}\left(\frac{l}{a}\right)+\cdots
\eea
where
\bea
A= \lim_{n\to 1} - \frac{\td}{\td n} C^1_1(n); \qquad B=-A
\eea
Note that, $C^1_1(n)$ can be computed using the relation
\bea
\frac{\langle \, \mathcal{T}(0)\widetilde{\mathcal{T}}(l)\, \mathbf{1} \, \rangle}{\langle \, \mathcal{T}(0)\widetilde{\mathcal{T}}(l) \, \rangle} = \langle \, \mathbf{1}\, \rangle_{\Omega,\text{cone}}
\eea
The left hand side (L.H.S.) of the above equation becomes
\bea
\text{L.H.S.}= 1+ C_1^1(n) \left(\frac{l}{a}\right)^{2-\Delta}
\eea
Computing the right hand side (R.H.S.) on the cone for sine-Gordon theory we can find $C^1_1$. The action for the sine-Gordon theory on the cone can be found to be
\begin{align}
\label{eq:bare_action_cone}
\mc{A}_c &= \int \td^2w \left[ \pd_w \phi \,\pd_{\bar{w}} \phi -\frac{1}{2}n^{\Delta -2} \left(\frac{w
	\bar w}{R^2}\right)^{\frac{(\Delta -2) (n-1)}{2 n}}\frac{\lambda_0}{\beta^2 a^2} \text{cos}(\beta \phi(w))\right]
\end{align}
Introducing the parameter, $ \delta = (\Delta-2)/2 $,
\bea
\text{R.H.S.}=\langle \, \mathbf{1}\, \rangle_{\Omega,\text{cone}} = 1 \ &+ & \frac{\lambda_0^2}{32 \pi^2 \Delta^2} \cross n^{4\delta} \cross \left(\frac{a^2}{R^2}\right)^{2\delta} (R^2)^{2+2\delta} \nonumber \\
&\cross & \int \frac{\td^2w_1}{R^2} \int \frac{\td^2w_2}{R^2} \left| \frac{w_1 w_2}{R^2}\right|^{\frac{2\delta(n-1)}{n}} \frac{1}{|w_1-w_2|^{2\Delta}} 
\eea
to $\mathcal{O}(\lambda^2)$. Near $\delta = 0$ one can expand in a power series in $\delta$. There are two possible sources for $\ln ({a\over R})$. One is from the pre-factor $({a^2\over R^2})^{2\delta}$ and the other is from the expansion  in powers of $\text{ln} \,|w_1w_2|$ of the integrand $\left| \frac{w_1 w_2}{R^2}\right|^{\frac{2\delta(n-1)}{n}}$. It is easy to see that the latter does not contribute anything because it is a UV finite integral. Thus one can ignore the factor $\left| \frac{w_1 w_2}{R^2}\right|^{\frac{2\delta(n-1)}{n}}$ in the integrand and the R.H.S. becomes
\be
\label{eq:Zc2n_delta0}
\langle \, \mathbf{1}\, \rangle_{\Omega,\text{cone}} \big |_{\delta\approx 0} = 1- \frac{\lambda_0^2}{64 }\cross \frac{n^2}{4} \left[ \left(\frac{a^2}{R^2}\right)^{2\delta} -\frac{R^2}{a^2} \right]
\ee
Again as before the quadratically divergent term in the above expression can be taken care of by introducing an appropriate cosmological constant term in the action. 

Finally comparing the R.H.S. with L.H.S. we find
\bea
C^1_1 (n)= \frac{\lambda^2}{64 }\cross \frac{n^2}{4}
\eea
and
\bea
B = \frac{\lambda^2}{64} \times \frac{1}{2}
\eea
Plugging the above expression in \eqref{eq:ee_twist} we find the entanglement entropy to be
\bea
\Delta S_{\pd A} =  \frac{\lambda_R^2 (\Delta -2)}{128}  \Log{\frac{l}{a}} + \mathcal{O}\left((\Delta-2)^2\right)
\eea
which matches with our previous result \eqref{EE}.  \\

%%%%%%%%%%%%%%%%%%%%%%%%%%%%%%%%%%%%%%%%%%%%%%%%
\section{\label{sec:norm}Normalization}
Here we normalize the bulk scalar field such that its boundary value which couples to the boundary
primary operator reproduces correct two point function in the boundary field theory. A detailed computation
is presented (also see \cite{Freedman:1998tz,Muck:1998rr}). \\

The (Euclidean) action of a massive scalar field on $AdS_{d+1}$  is given by
\bea
\mathcal{A}= \frac{1}{2} \int \td^{d+1} x \, \sqrt{g} \left[ g^{\mu\nu} \pd_\mu \Phi \pd_\nu \Phi + m^2 \Phi^2 \right]
\eea
The equation of motion is
\bea
\frac{1}{\sqrt{g}} \pd_\mu \left( \sqrt{g} g^{\mu\nu} \pd_\nu \Phi \right) - m^2 \Phi =0
\eea

which in Poincare patch yields
\be \label{eq:wave}
 z^{d+1} \frac{\pd}{\pd z} \left[z^{-d+1}\frac{\pd}{\pd z} \Phi(z,\vec{x})  \right]+ z^2 \frac{\pd^2}{\pd \vec{x}^2} \Phi(z,\vec{x})-m^2 \Phi(z,\vec{x})=0
\ee

The general solution of the above equation is

\be 
\label{eq:fourier}
\Phi(z,\vec{x}) = \frac{1}{(2\pi)^{d/2}}\int \td \vec{k}\,\Exp{i \vec{k}\cdot\vec{x}}\, \Phi(z,\vec{k})
\ee
with
\be
\label{eq:gen_sol}
\Phi(z,\vec{k}) = c_1 z^{d/2} I_\nu (i k z) + c_2 z^{d/2} K_\nu (i k z)
\ee
where $I_\nu (i k z), K_\nu (i k z)$ are modified Bessel functions with index
\be 
 \nu = \left[\frac{d^2}{4}+m^2\right]^{1/2}.
\ee
Inserting \eqref{eq:fourier} in the action and performing integration by parts one can find
\bea
 && ~~~\mathcal{A} = \frac{1}{2} \int \td\vec{k} ~\td \vec{k'} ~\delta (k+k') \left[ z^{-d+1} \Phi(z,\vec{k}) \pd_z \Phi(z,\vec{k'}) \right]_a^\infty \cr 
 && - \frac{1}{2} \int \td\vec{k} ~\td \vec{k'} ~\delta (k+k') \frac{\Phi(z,\vec{k})}{z^{d+1}} \left[z^{d+1} \pd_z\left(z^{-d+1}\pd_z \Phi(z,\vec{k'})\right)-(k^2 z^2+m^2) \Phi(z,\vec{k'}) \right] \nonumber
\eea
The last term in the above expression vanishes when the equation of motion \eqref{eq:wave} is satisfied. 
Thus the on shell action becomes  
\be
\mathcal{A} = \frac{1}{2} \int \td\vec{k} ~\td \vec{k'} ~\delta (k+k') \left[ z^{-d+1} \Phi(z,\vec{k}) \pd_z \Phi(z,\vec{k'}) \right]_a^\infty
\ee
Let the solution to \eqref{eq:wave} be of the following form
\be 
 \Phi(z,\vec{k}) = f^a(z,\vec{k}) \lambda_b (\vec{k})
\ee
such that
\be 
\label{eq:bdy_condition}
 \lim_{z \to a} f^a(z,\vec{k}) =1, \qquad \lim_{z \to \infty} f^a(z,\vec{k}) =0.
\ee
With these boundary conditions \eqref{eq:bdy_condition} the solution to the equation of motion \eqref{eq:gen_sol} becomes
\be 
 f^a(z,\vec{k}) = \left( \frac{z}{a}\right)^{d/2} \frac{K_\nu(k z)}{K_\nu(k a)}
\ee
as the modified Bessel function $K_\nu (kz)$ vanishes as $z\to \infty$.

From AdS/CFT correspondence 
\be 
 \exp (-\mathcal{A}_{AdS}) \equiv \left\bra \exp \left(\int \td^d x J(\vec{x})\mathcal{O}(\vec{x})\right) \right\ket = Z[J]
\ee
where
\be 
 J(\vec{x}) = \lim_{a\to 0} a^{\nu-\frac{d}{2}}\lambda_b(\vec{x}).
\ee
In momentum space the two-point function of a primary operator is given by
\be
 \bra \mathcal{O}(\vec{k})\mathcal{O}(\vec{k'})\ket =\lim_{a\to 0} a^{-2\nu+d} \left. \frac{1}{Z[0]} \left( \frac{\delta}{\delta \lambda_b(k)} \right)\left( \frac{\delta}{\delta \lambda_b(k')} \right)Z[\lambda_b] \right|_{\lambda_b=0}
\ee
where $Z[\lambda_b]$ is the partition function with source $\lambda_b$.

Therefore from bulk, the boundary two-point function will be
\be
\label{eq:2pt_func}
 \bra \mathcal{O}(\vec{k})\mathcal{O}(\vec{k'})\ket =\lim_{a\to 0}- a^{-2\nu+d} a^{-d+1} \delta(\vec{k}+\vec{k'}) \lim_{z\to a} \pd_z f^a(z, \vec{k})
\ee
As the behavior of $K_\nu (kz)$ near $z=0$, is given by
\be
 K_\nu (kz) = 2^{\nu-1}\Gamma[\nu] (kz)^{-\nu}[1+\cdots]-2^{-\nu-1} \frac{\Gamma(1-\nu)}{\nu} (kz)^\nu [1+\cdots] \ee
where the terms `$\cdots$' are positive powers of $(kz)^2$, and 
\be 
 \pd_z K_\nu(kz)=\frac{\nu}{z}K_\nu(kz)-kK_{\nu+1}(kz)
\ee
we find
\bea
\bra \mathcal{O}(\vec{k})\mathcal{O}(\vec{k'})\ket
 = -  \delta(\vec{k}+\vec{k'}) k^{2\nu}2^{-2\nu}\frac{\Gamma(1-\nu)}{\Gamma(1+\nu)} (2\nu)+\cdots
\eea
to the leading order in the limit $a\to 0$.
Taking Fourier transform of the above expression and using $\nu = \Delta -\frac{d}{2}$ one gets
\be
 \bra \mathcal{O}(\vec{x})\mathcal{O}(\vec{y})\ket =  \frac{2\nu \kappa}{|\vec{x}-\vec{y}|^{2\Delta}}
\ee
with
\be 
\kappa= \frac{\Gamma(\Delta)}{\pi^{d/2}\Gamma(\nu)}.
\ee
For $d=2$ the holographic 2d conformal correlation function becomes,
\be 
 \bra \mathcal{O}(\vec{x})\mathcal{O}(\vec{y})\ket =  \frac{2}{\pi} (\Delta-1)^2 \frac{1}{|\vec{x}-\vec{y}|^{2\Delta}}.
\ee

To recover our conventional result \eqref{2pt-fn} for the ``sine-Gordon operator'' with 
$\Delta = \beta^2/4 \pi$, we need to normalize the field in such a way so that the interaction 
term in the partition function $Z[\lambda_0]$ becomes
\be 
 \label{eq:field_norm} 
 \frac{2}{\sqrt{\pi}}\int \lambda_b \mathcal{O}
\ee

We use the above normalization \eqref{eq:field_norm} for the scalar field in our bulk computation of EE.

\bibliographystyle{bibstyle}
\bibliography{sine-gordon}
\end{document}